# Thermodynamic Mechanism of Life and Aging

Marko Popovic

Department of Chemistry and Biochemistry, Brigham Young University, Provo, UT 84602, USA

*"Life is a series of natural and spontaneous **changes**."*
*Lao Tzu, VI century B.C.*

**Abstract:** Life is a complex biological phenomenon represented by numerous chemical, physical and biological processes performed by a biothermodynamic system/cell/organism. Both living organisms and inanimate objects are subject to aging, a biological and physicochemical process characterized by changes in biological and thermodynamic state. Thus, the same physical laws govern processes in both animate and inanimate matter. All life processes lead to change of an organism's state. The change of biological and thermodynamic state of an organism in time underlies all of three kinds of aging (chronological, biological and thermodynamic). Life and aging of an organism both start at the moment of fertilization and continue through entire lifespan. Fertilization represents formation of a new organism. The new organism represents a new thermodynamic system. From the very beginning, it changes its state by changing thermodynamic parameters. The change of thermodynamic parameters is observed as aging and can be related to change in entropy. Entropy is thus the parameter that is related to all others and describes aging in the best manner. In the beginning, entropy change appears as a consequence of accumulation of matter (growth). Later, decomposition and configurational changes dominate, as a consequence of various chemical reactions (free radical, decomposition, fragmentation, accumulation of lipofuscin-like substances…).

**Introduction**

"*What is life?*" is a question first raised by Schrodinger [1944], who wanted to show the necessity that life processes should obey the laws of physics and chemistry. However, there is no general agreement in the scientific community about the answer, even today. An even more controversial question is "what is aging", except for the general agreement that aging is an irreversible process closely related to life and time flow. There are many hypotheses about the cause of aging.

The literature describes the seven characteristics in common to all living organisms. Living organisms (1) are built of structures called cells, (2) grow and change, (3) have a complex chemistry, (4) maintain homeostasis, (5) respond to their environment, (6) reproduce and have offspring, and (7) pass their traits onto their offspring.

According to the first characteristic, all living organisms have cellular structure. A cell corresponds to the description of a thermodynamic system (Figure 1). It has a certain amount of matter, clearly separated by boundaries from its surrounding, with which it exchanges matter and energy. Thus, a cell can be considered as an open thermodynamic system, as was shown by Von Bertalanffy [1950].

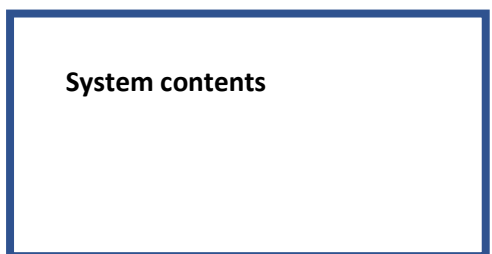
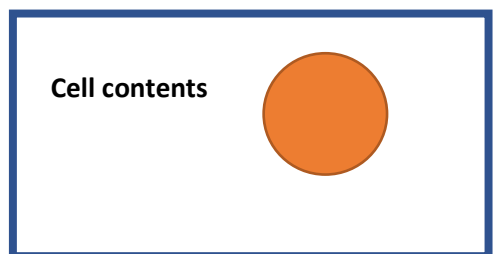

**Figure 1:** Cells structures corresponds to a structure of thermodynamic system.

A thermodynamic system or a cell is characterized by state parameters (e.g. volume *V*, pressure *p*, temperature *T*, mass *m*, amount of matter *n*, enthalpy *H*, entropy *S*, internal energy *U*, Gibbs free energy *G*…). Any change of state parameters represents a process.  It is obvious that life and aging processes lead to change of state. Indeed, during life (and aging), there are consecutive irreversible changes in state parameters of a system/cell/organism.

(State 1) -> (State 2) -> (State 3) -> … -> (State N) -> … -> (Death) -> (Decomposition)  (1)

---------------------------------------------------------------------------------------------------------------> Time

If a cell/organism is considered as a thermodynamic system, then each state is characterized by certain state parameters. State 1 is characterized by mass $m_1$, volume $V_1$, internal energy $U_1$, enthalpy $H_1$, entropy $S_1$, and Gibbs free energy $G_1$, amount of information $I_1$. During aging, there is a process that changes the state parameters. Life occurs at approximately constant temperature (a very narrow temperature range) and constant pressure. Thus, aging is a process that is isobaric and isothermal. Variable parameters are volume (which we macroscopically observed as growth). The increase in volume is caused by an accumulation of mass. Because of the change in mass, there is a change in enthalpy, internal energy, entropy, Gibbs free energy, information… Therefore, state 2 is characterized by properties $m_2$, $V_2$, $U_2$, $H_2$, $S_2$, $G_2$, $I_2$… State of a living organism changes during aging. Aging is a complex phenomenon, which includes physiological changes (loss of function), cytological changes (accumulation of lipofuscin, telomere shortening, cell damage…) and histological changes (loss of parenchymal cells and their replacement with connective tissue).

 Almost all of 300 existing theories of aging are based on certain observations in regard to morphological and biochemical changes which occur during aging. However, none of them explains the physical reason that causes the aging process.  No single theory is generally accepted [Harman 1998].  The popular free radical theory of aging [Harman 1992; Harman 1998; Yin 2005; Harman 1956; Harman 1980] is based on the chemical nature and ubiquitous presence of free radicals [Ashok 1999]. It suggests that aging changes are caused by free radical reactions [Harman 1992]. The oxidative damage theory of aging once seemed almost proved. But recently it was criticized. So, an attempt was made to modify it. According to the proposed neo-oxidative damage theory, oxidative damage is a major contributor to aging, but $O_2^-$ and $H_2O_2$ are not the main causes of this damage [Gems 2009].

The evolutionary theory of aging is currently the most popular aging theory. Kirkwood [2000] reports that specific genes selected to promote aging are unlikely to exist. According to it, aging is not programmed but results largely from accumulation of somatic damage, owing to limited investments in maintenance and repair. However, the genetic determination of aging [Johnson 1987] suggests that development, reproduction, and life span are each under independent genetic control.

There is experimental evidence in support of the telomere-telomerase hypothesis of aging [Holt 1996]. Damage to DNA is the centerpiece of this theory. Recently telomere shortening has been described to be associated with DNA damage [Ahmed 2001]. Some cells do not possess any genetic material (i.e. erythrocytes, platelets…) but they are still subject of the aging process. In that case genetic determination of aging is pointless. Inanimate mater is also subject of the aging process. Moreover, aging process starts immediately after fertilization when nucleic acids are not damaged.

Other theories of aging are based on some specific observations. Cellular manifestations of aging are most pronounced in postmitotic cells. Many mitochondria undergo enlargement and structural disorganization while lysosomes, which are normally responsible for mitochondrial turnover, gradually accumulate an undegradable, polymeric, autofluorescent material called lipofuscin [Brunk 2002].

Some authors offered a consideration of biological aging as the result of failure of homeostatic systems [Adelman 1980; Gerhard 1993]. Homeostatic systems have a role in the maintenance of the balance inside a biological system. Their failure leads to changes which characterize the increase of entropy inside the cell. However, the aging process begins long before the failure can appear.

Hayflick [2007plos; 2007pnas; 1998; 2016] suggests that potential longevity is determined by the energetics of all molecules present at and after the time of reproductive maturation. The relation between increase of entropy and aging was studied, both theoretically and experimentally by Hayflick [2007plos; 2007pnas; 1998; 2016], Toussaint [1995; 2000; 2002; 1998; 1991], and others [Popovic 2012; Silva 2008; Shamir 2009; Popovic 2012; Schneider 2005]. Some authors are explicit, reporting that the entropy of the *C. elegans* pharynx tissues increases as the animal ages [Shamir 2009]. Silva and Annamalai [2008] reported an increase of entropy of the organism during aging, 11.404 kJ/K per kg of body mass. The rate of entropy generation was found to be three times higher on infants (because of growth) then on the elderly. This is in accordance with the reports of Toussaint [1995; 2000; 2002; 1998; 1991] that entropy generation of cells decreases during their life. This can be explained by the fact that in youth increase in entropy is mostly caused by growth, while during senescence entropy increase is the result of various degenerative processes. Thermodynamic studies of biology of aging are extended by the ongoing studies on longevity extension by caloric restriction (input of substance restriction) in rodents and primates [Roth 2002; Weindruch 2001; Mobbs 2001; Poehlman 2001].

Aging is also a physical and a chemical process (which is manifested in changes in configuration and conformation of molecules), which are a consequence of various factors (e.g. ROS theory, entropy explanation of aging…). This leads to a change of state of a thermodynamic system/cell/organism.

Aging is a phenomenon observed in all living organisms and is also present in non-live matter. Thus, it can be concluded that aging is a universal natural phenomenon.

In this paper it is hypothesized that the aging process is a spontaneous thermodynamic process performed by a living organism, an open thermodynamic system out from equilibrium with a property of growth. During this process there is a change of not only one, but several parameters.

To consider separately aging of living systems resembles the vitalistic approach. However, biological systems remain the main subject of most aging studies. Aging is a process of state change of both living systems and objects. The causes of aging are in common to both living and non-living systems. Fundamentally, both kinds of aging can be in the most general sense be related to change of state of a system, caused by the process that leads to the change in state parameters. The difference in aging of living and non-living systems is that in living systems there is a positive change in parameters ($\Delta V > 0$, $\Delta m > 0$, $\Delta S > 0$, $\Delta U > 0$). However, in non-living systems the change in volume, mass, amount of matter, internal energy and entropy can be less than, equal to or greater than zero.

**Theoretical consideration**

It seems obvious that life begins with the process of fertilization. The fertilization process represents the moment of formation of a thermodynamic system (zygote, future organism) from two components. It seems obvious that, from fertilization, through cell division growth and aging occur. Here a question appears, are growth and aging two distinct parallel processes or is growth an integral part of the aging process? In this process there is an increase in mass, volume and amount of matter in the system. Because of mass accumulation, there is an accumulation of entropy, enthalpy and internal energy. Because of the entropy increase, Gibbs free energy is negative. Thus, growth is a spontaneous process caused by accumulation of matter. Immediately after the formation of a zygote, the changes in these parameters lead to change of state. If change of state of the system is defined as aging, then the aging process begins from the moment when an organism is formed. Indeed, aging of embryos can be monitored chronologically through days and months, but also thermodynamically as irreversible, consecutive and continuous changes of state. Aging of an organism begins with its formation and ends with its death. After death, aging continues as aging of non-living systems or objects.

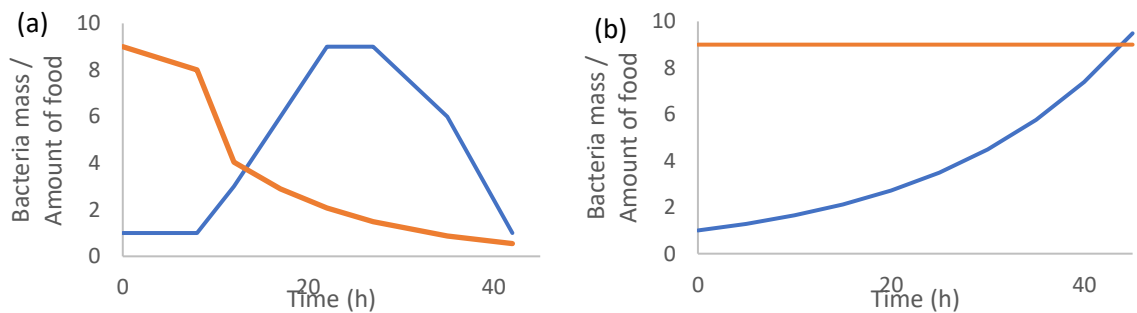

**Figure 2**: Continuous change of biological and thermodynamic state: (a) An organism situated in a closed system (Petri dish) trough accumulation of matter increases its mass. By increasing its mass, it increases its entropy, enthalpy and internal energy. (b) An organism situated in an open system also exponentially increases its mass and entropy.

It can be concluded that during the life process and aging, an organism changes more than one thermodynamic parameter. Thus, aging represents a complex thermodynamic process. In that process, state of the organism changes irreversibly, consecutively and continuously from one state to another. Each state is characterized by its own state parameters. It is hard to limit the analysis of aging to change in just one parameter (e.g. telomere shortening, accumulation of lipofuscin…). As shown above an organism changes more than a few parameters during state transitions, so the changes in all state parameters must be considered, while analyzing the aging process. The reason is that a change in one of the parameters leads to changes in others. It seems that the driving force of life is also the reason of the aging process. Therefore, aging begins from fertilization and ceases with death. Therefore, it can be concluded that life and aging are two sides of the same coin.

People intuitively understand the aging process chronologically. Many consider time flow to be the cause of aging. However, different organisms age at different rates. Even though the correlation between time and aging obviously exists, different organisms do not change their states in the same way during a same time. Macroscopically, this means that people age at different rates. The intensity of changes that characterize aging can be different in people of same chronological age. The next obvious example of the lack of correlation between aging and time is premature senescence. This leads to the conclusion that time flow itself is not the main cause of aging.

It is a fact that aging does not begin with lipofuscin accumulation, oxidative damage or telomere shortening (which appears later in life), but with the first cell division and continues during entire lifespan. Therefore, changes in state of an organism appear since the very beginning – fertilization and zygote formation. Indeed, immediately after fertilization, a cell begins to grow and after that divide. This leads to change in parameters of states and states of the system (organism), which is characteristic of aging. This leads to the conclusion that, even though changes in structure and function characterize aging and significantly contribute to the change of state, they are not the only driving force of the aging process. Even though structural changes of molecules significantly contribute to the aging process, they also appear later in life and cannot be responsible for the change of state of an organism during embryonic development (especially at the very beginning).

If aging represents a change of state and a cell represents a thermodynamic system, then the fundamental driving force for the change of state is accumulation of matter (mass and energy) and accumulation of errors. This accumulation of matter occurs in parallel with time flow and cell growth either in volume (parenchymal cells) or by division. In both cases there is an increase in mass and volume of an organism, which leads to state transition. Mass accumulation, except for the volume increase, leads to increase in enthalpy and entropy.

Aging is a spontaneous and irreversible process. From this it can be concluded that the free energy change is negative $\Delta G < 0$. It is obvious that, according to the Gibbs equation, the entropy contribution during aging is greater than the enthalpy contribution.

$$\Delta G = \Delta H - T\Delta S \tag{2}$$

From this equation it follows that the entropy change during aging is greater than the enthalpy change.

The change in mass is positive, as a consequence of growth, and specific entropy of all substances is greater or equal to zero because of the third law of thermodynamics. Therefore, the change in entropy of the organism must be positive. Change in entropy can be related to changes in all other parameters related to aging. Entropy change is related to mass change through the equation

$$\Delta S = s\Delta m \tag{3}$$

$s$ is specific entropy or entropy per unit mass [Balmer 2010]. Entropy change can also be related to change in volume, by using the definition of density: $\rho = m / V$.

$$\Delta S = s\rho\Delta V \tag{4}$$

Internal energy change can also be related to entropy change. Internal energy change is related to change in entropy by combining equation (3) with $\Delta U = u\, \Delta m$, where $u$ is specific internal energy or energy per unit mass.

$$\Delta S = (s/u)\Delta U \tag{5}$$

if $u$ and $s$ are constant during the process. Similarly, enthalpy change can be related to entropy change by combining equation (3) with $\Delta H = h\,\Delta m$, where $h$ is specific enthalpy or enthalpy per unit mass

$$\Delta S = (s/h)\Delta H \tag{6}$$

if *h* and *s* are constant during the process. Gibbs free energy can also be related to entropy change, by combining equations (2) and (6) and rearranging, which gives

$$\Delta S = [(h/s) - T]^{-1} \Delta G \tag{7}$$

if *h, s* and *T* are constant during the process.

Molecular changes characteristic of aging can also be related to entropy change, according to Hayflick [2016]. This is because entropy is related to the number of microstates available to a system *W* through the Boltzmann equation *S* = *k* ln*W*, where *k* is the Boltzmann constant [Dugdale 1996; Atkins 2008]. Entropy change is thus

$$\Delta S = k \ln(W_2/W_1) \tag{8}$$

$W_1$ and $W_2$ are the number of microstates available to a system in states 1 and 2, respectively. Therefore, entropy change quantifies molecular-level changes in a system.

It seems obvious that entropy change describes changes of state characteristic of aging in the best way. This entropy change in early phases is a consequence of change of parameters caused by matter accumulation. Later in life growth and matter accumulation cease. The entropy change in the adult and senescent phase is caused mostly by changes in structure of live matter and decomposition processes (molecules, cells, tissue). However, accumulation does not cease completely, since "erroneous" molecules (lipofuscin) continues to accumulate. Another advantage of using entropy to quantify aging is that it is the subject of the second law of thermodynamics. This gives additional constraints for analysis and automatically removes models that are not physical meaningful. Furthermore, all thermodynamic systems behave according to the laws of thermodynamics, which is well established and experimentally validated, making the analysis reliable.

Apoptosis cannot be the main mechanism of aging, because organism begins to age right after fertilization, long time before apoptosis can appear.

**Conclusions**

Aging represents a universal irreversible complex thermodynamic phenomenon, characteristic of both living and nonliving matter. Basically, aging represents an irreversible consecutive change of states characterized by change of many thermodynamic and biological parameters. In order to simplify the analysis of the aging process, it is suggested to use entropy, since changes in the other parameters are included into entropy change. Use of other parameters, such as internal energy, volume, mass, various structural changes or information changes, is incomplete since they give a partial picture and none of them includes all the others. All biological, functional (physiological) and morphological, as well as molecular changes lead to change of entropy of an organism. This is one more reason why the change in entropy is the best choice for describing the aging process. Multi-causal continuous change in entropy content of an organism represents the driving force of life and aging.